# Theoretical Principles of Single-Molecule Electronics: A Chemical and Mesoscopic View

Yongqiang Xue[1,*], Mark A. Ratner[2]

[1]College of Nanoscale Science and Engineering, University at Albany, State University of New York, Albany, NY 12203

[2]Department of Chemistry and Materials Research Center, Northwestern University, Evanston, IL 60208

[*]Author to whom correspondence should be addressed. Email: yxue@uamail.albany.edu

**ABSTRACT:** Exploring the use of individual molecules as active components in electronic devices has been at the forefront of nanoelectronics research in recent years. Compared to semiconductor microelectronics, modeling transport in single-molecule devices is much more difficult due to the necessity of including the effects of the device electronic structure and the interface to the external contacts at the microscopic level. Theoretical formulation of the problem therefore requires integrating the knowledge base in surface science, electronic structure theory, quantum transport and device modeling into a single unified framework starting from the first-principles. In this paper, we introduce the theoretical framework for modeling single-molecule electronics and present a simple conceptual picture for interpreting the results of numerical computation. We model the device using a self-consistent matrix Green's function method that combines Non-Equilibrium Green's function theory of quantum transport with atomic-scale description of the device electronic structure. We view the single-molecule device as "heterostructures" composed of chemically well-defined atomic groups, and analyze the device characteristics in terms of the charge and potential response of these atomic groups to perturbation induced by the metal-molecule coupling and the applied bias voltage. We demonstrate the power of this approach using as examples devices formed by attaching benzene-based molecules of different size and internal structure to the gold electrodes through sulfur end atoms.

## 1. Introduction

Assembling materials and device structures by arranging atoms the way we want them has been an elusive goal for scientists and engineers ever since the discovery of individual atoms/molecules as the basic elements of all chemical and biological processes. Richard Feynman told us that there is no physical limitation to such atom-by-atom manipulations [1], and there is no physical limitation to information processing from the quantum mechanical laws governing the microscopic atomic world either [2]. Aviram and Ratner proposed the first molecular electronic device in 1974 [3] by suggesting that a molecule with a donor-spacer-acceptor structure may behave as a diode when connected to the electrodes. There has been a long history of investigating the electrical behavior of molecular and organic materials and exploring their applications in enabling device technologies, including microelectronics [4] and chemical- and bio-sensors [5]. But the laboratory realization of devices functioning on the single-molecule basis has to wait for the advent of scanning probe microscopy techniques in the 1980s [6], which, for the first time, allow scientists and engineers to image, manipulate and fabricate materials and device structures on the basis of individual atoms and molecules [7].

Today, the relentless driving toward device miniaturization by the semiconductor industry during the past three decades has pushed silicon-based microelectronics down to the nanometer range [8,9], following the well-known design scaling rules derived from semi-classical semiconductor transport equations [10]. For semiconductor technology, research on single-molecule electronics is essentially a rehearsal of what we may expect as silicon-based microelectronics shrink to the atomic and molecular regime. The recent surge of activity on molecular electronics [11] thus reflects the convergence of the trends of top-down device miniaturization through the lithographic approach [8,9] and bottom-up device fabrication through the atom-engineering and self-assembly approaches made possible by the advancement of nanotechnology [7,12]. Although it remains unclear about the optimal materials, device structure and circuit of integrated molecular nanoelectronics and its performance limit, the opportunities that molecular electronics present on the prospect of device design through atomic-level engineering have undeniably far-reaching impacts on both nanostructured devices in general and for the further down-scaling of silicon microelectronics into the atomic regime in particular.

The building block of molecular electronics is the two-terminal single-molecule device (Figure 1), where the molecule is attached to the surfaces of predefined electrodes through appropriate end groups or "alligator clips" [11]. The electrodes may not be metals and/or perfect single-crystalline materials, where atomic-scale structures may lie on top of the semi-infinite substrates. In the case of semiconductor electrodes, surface reconstruction may also occur. Although early proposals on molecular electronics have focused on the intrinsic properties of the molecules, it is now widely recognized that the nature of the electrode-molecule interface plays a significant, sometimes dominant, role in interpreting the measured transport characteristics of the molecular junctions. Detailed understanding of the electronic and transport properties of single-molecules contacted by electrodes is thus crucial for the development of molecular electronics. A number of experimental strategies have been attempted to probe electron transport through single-molecules in such electrode-molecule-electrode heterojunctions, using, e.g., mechanical break junctions, scanning nanoprobe microscopes and crossed nanowires [11]. Concurrently, a variety of theoretical approaches have been applied to model electron transport in such devices, ranging from semi-empirical to *ab initio* methods.

Like its experimental counterpart, much of the modern theory of molecular electronics can trace its origin to the theoretical development in the field of scanning tunneling microscopy and spectroscopy [13,14]. These early developments highlight two key ingredients of molecular electronics: (1) atomic-scale surface physics of the electrical contacts and (2) chemistry of the molecule and surface-molecule interaction. The addition of third ingredient, i.e., quantum transport theory, turns the theory of molecular electronics into a well-defined subfield of nanoscience and nanotechnology. In this work, we take the view that current research activities on single-molecule electronics are the natural continuation and scaling limit (in all three dimensions) of researches on semiconductor contacts [15,16] and mesoscopic quantum semiconductor devices [17], the hallmarks of which are the contact barrier for charge injection, quantum confinement, band "engineering" and phase coherent carrier transport from one contact of the sample to another contact.

In view of the rapid progress in device fabrication, transport measurement and numerical modeling of molecular scale devices and the variety of experimental and theoretical approaches adopted, a critical examination of the fundamental physical processes involved in single-molecule electronics from a single and unifying theoretical framework, followed by detailed numerical examples of selected aspects of device design through bottom-up atom engineering and their rationalization through the use of several central concepts, will help to explore the full potential and/or limitation of molecular electronics, which is the purpose of the present introductory work. The rest of the paper is organized as follows: We present the general theoretical model in section II. We discuss the essential physics of single-molecule electronics using as examples the prototypical devices formed by attaching benzene-based molecules to the gold electrodes through sulfur end atoms. We separate the discussion of device physics of single-molecule electronics into device at equilibrium in section III and device at non-equilibrium in section IV. We conclude in section V.

## 2. Theoretical Model

### 2.1. MESOSCOPIC VIEW OF SINGLE-MOLECULE ELECTRONICS

Traditional quantum mechanical theory of electron transport typically views the electric field as a cause and the current flow as a response, which leads to essentially a generalized theory of polarizability, e.g., Kubo's linear-response theory [18]. Such theory ignores explicit interface of the sample to the rest of the electrical circuit. The development of mesoscopic physics in the past two decades [17] has introduced an alternative view, i.e., transport is a result of the carrier flow incident on the sample boundary, and the field distribution within the sample results from the self-consistent pileup of non-equilibrium charge carriers. For small conduction systems where the carriers have a totally quantum mechanical coherent history inside the sample, this leads to the celebrated Landauer approach, where transport is viewed as a transmission process similar to electromagnetic wave propagation in an optical waveguide [19].

The transmission view is essentially a generalization of the circuit theory with the contacts being treated as known source of carriers, similar to a voltage or a current node in a classical electrical circuit. Correspondingly, the finite sample does not have an unique conductance: its conductance depends on how the current source is introduced. The circuit plot of the two-terminal single-molecule device in Figure 1(a) is a simplified view of realistic experimental configurations. The electrodes are usually microscopic large but macroscopic small contact pads connected to the macroscopically large battery through leads, which are transmission lines with distributed (classical) resistance, capacitance and inductance etc. Since the electrodes have much larger dimension than the molecules, the molecular device can be considered as forming an adiabatic, widening contact to the electrodes. The widening contact, due to the geometrical aspect of the molecule-electrode interface, makes it a good approximation to consider the electrodes as thermal electron reservoirs in typical molecular electronics measurement [19, 20].

Here the reservoirs are the electronic analog of the radiative blackbody: they absorb incident carriers without reflection and they emit carriers with a (fixed) thermal equilibrium distribution. They are characterized by an effective Fermi-level, or effective electrochemical potential. The conductance of the molecular device can only be calculated after specifying the location where the Fermi-levels are determined. It is clear that the electrochemical potential difference relevant for current calculation may be different from the voltage measured from the battery, which includes the voltage drops across the series resistance due to the external leads. In the case of semiconductor electrodes, it also includes any potential drop across the space-charge region due to depleted dopants and/or band discontinuity away from the molecule-electrode interface [21].

## 2.2. THE "EXTENDED MOLECULE" CONCEPT IN SINGLE-MOLECULE ELECTRONICS

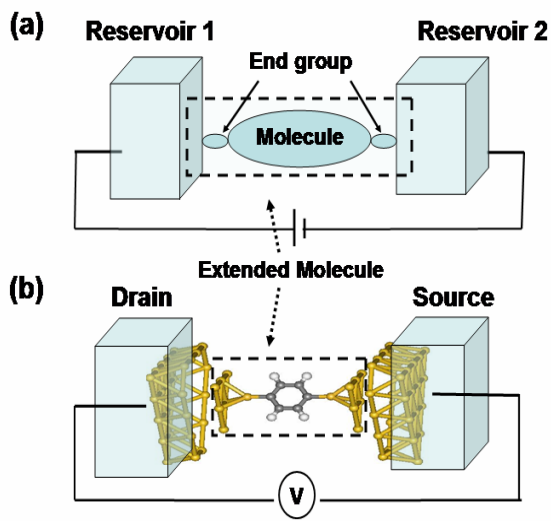

**FIGURE 1.** (a) Schematic illustration of a two-terminal single-molecule device. (b) Atomic structure of the "extended molecule" and contact region for device formed by attaching phenyl dithiol molecule to two gold surfaces through the end sulfur atoms. The surface atoms included into the "extended molecule" have been moved out of the surface to aide visualization.

The investigation of electron transport through single-molecules starts with the specification of the reservoir and the active device region. We assume that the two electrodes are sufficiently far away from each other so that direct tunneling between them is negligible compared to the contribution mediated by the molecular bridge. The bare bi-electrode junction (without the molecular insertion) is thus a double-plate capacitor. The electron density and electrostatic potential distribution in such bi-electrode junction can be calculated using standard electronic structure methods under both zero and non-zero biasing conditions, since each electrode remains in thermal equilibrium while the bias voltage is maintained through the external circuit (i.e., the battery). In general, the property of the bi-electrode capacitor depends on the atomic-scale detail of the contacts involved [22].

Introducing molecules into the junction short-circuits the bi-electrode capacitor and can lead to arbitrarily complicated charge and potential distributions in the single-molecule device depending on the internal and end-group structures of the molecules. Since the interaction between the molecule and the electrodes are in general not weak, this needs to be considered treating the electrode-molecule-electrode heterojunction as a unified coherent quantum system. Given the charge and potential distributions of the bi-electrode capacitor, an active device region can be conveniently defined as the part of the molecular junction where the charge and potential distributions differ from those of the bi-electrode junction for the given bias. Such region includes necessarily both the molecule itself and finite number of atoms on the surface of each electrode involved, which is called "extended molecule" (EM) from here on. Correspondingly, the rest of the molecular junction (with the "extended molecule" excluded) comprises the contact region which can be treated as electron reservoirs with fixed electrochemical potentials. The "extended molecule" is charge neutral, since otherwise the electronic states in the contact region will be perturbed by the Coulomb field due to the net charge within EM.

The "extended molecule" concept plays a central role in analyzing the device physics of single-molecule electronics: (1) From the electronic structure point of view, the effect of metal-molecule interaction is "absorbed" entirely within EM since the electronic states within the contact region are identical to those of the bi-electrode junction and are not affected by the molecular adsorption. Since EM contains an integer number of electrons, this allows us to analyze metal-molecule interaction using a qualitative molecular orbital picture [23], where the surface part of the EM comprises a chemically well-defined atomic group similar to any other functional groups within the molecule itself. (2) From the electronic transport point of view, the "extended molecule" presents an atomic-scale analog of semiconductor quantum heterostructures, where the hetero-interface can be introduced either at the metal-molecule interface or between different functional groups inside the molecule. Current transport through the single-molecule device is determined by the non-equilibrium state of EM through the non-equilibrium charge pileup induced by carriers injected from the reservoirs. The intrinsic inhomogeneous structure of EM may lead to strong local spatial variation of current and electric field, which can be modulated separately by chemical modification of the metal-molecule interface and the internal structure of the molecule and lead to potentially interesting nonlinear transport characteristics [24, 25].

## 2.3. REAL-SPACE NON-EQUILIBRIUM GREEN'S FUNCTION APPROACH TO SCATEERING THEORY OF SINGLE-MOLECULE ELECTRONICS

The "extended molecule" is a physically well-defined concept. The numerical computation will involve approximations. In particular, accurate first-principles techniques are indispensable describing the electronic processes within the finite "extended molecule" where all the interesting physics occur, while the contact region can be described using lower-order theory without affecting the extraction of essential physics involved. We base our analysis of single-molecule electronics here on a self-consistent matrix Green's function method (SCMGF) which combines the Non-Equilibrium Green's Function (NEGF) theory of quantum transport [26,27] with atomic-scale description of the device electronic structure and uses the technique of expansion in finite local atomic-orbital basis sets [20]. Within the phase coherent transport regime and an effective single-particle theory of the interacting electron systems, the NEGF theory is formally equivalent to the scattering theory of mesoscopic transport [28]. By replacing the quasi-particle exchange-correlation self-energy $\Sigma_{xc}$ [29] with the exchange-correlation potential $V_{xc}$ within the Kohn-Sham (KS) density functional theory (DFT) of interacting electron system [30], the well-established technique of self-consistent field theory of molecular and surface electronic structure can then be utilized for transport calculations [20]. In the field of molecular electronics and related scanning tunneling microscopy research, wave function-based methods have also been used widely within the scattering theory of mesoscopic transport, often in conjunction with the jellium approximation of metallic electrodes [31, 32]. The main advantage of using SCMGF method here is the ease to include the contact effects, the automatic inclusion of the Pauli Principle of particle symmetry, the simultaneous treatment of the electronic structure and the non-equilibrium transport effects within the same theoretical framework through the connection to the Green's function based electronic structure method [33] and the connection to the rigorous many-body quasi-particle theory of interacting electron and electron-phonon systems [30]. The details of the SCMGF theory have been described elsewhere [20, 34], so we only give a brief summary here to show how the physical observables are computed. Note that similar methods have been developed independently by several other research groups [35].

Expanding the wave function and Green's function of the finite active device region in terms of local atomic-type orbital basis $\phi_i(\mathbf{r})$ in real space as $\psi_i(\mathbf{r}) = \sum_j c_{ij} \phi_j(\mathbf{r})$ and $G_{Dev}^{r(<)}(\mathbf{r},\mathbf{r}',E) = \sum_{ij} G_{ij}^{r(<)} \phi_i(\mathbf{r}) \phi_j(\mathbf{r}')$ respectively, the SCMGF method leads to the solution of the following Dyson and Keldysh-Kadanoff-Baym (KKB) equations for the Green's functions in the matrix format:

$$G_{Dev}^r(E) = (E^+ S_{Dev} - H_{Dev} - \Sigma_L^r - \Sigma_R^r)^{-1},$$
$$G_{Dev}^<(E) = G_{Dev}^r(E)[i\Gamma_L f(E-\mu_L) + i\Gamma_R f(E-\mu_R)]G_{Dev}^a(E),$$
$$\Sigma_{L(R)} = (E^+ S_{ML(R)} - H_{ML(R)})G_{LL(RR)}^{0,r}(E)(E^+ S_{L(R)M} - H_{L(R)M}),$$
$$\Gamma_{L(R)} = i(\Sigma_{L(R)}^+ - \Sigma_{L(R)}).$$
(1)

where $S_{Dev}$ is the overlap matrix and $G_{LL(RR)}^0$ is the surface Green's function of the left (right) contact. $H_{Dev}$ is the self-consistent Hamiltonian matrix of the active device region (the "extended molecule"), which includes the Hamiltonian describing the "extended molecule" itself, the external field describing the applied bias voltage and the electrostatic potential due to the charge distribution outside the active device region and any self-consistent charging effect due to charge transfer and field-induced charge redistribution [20]. The self-consistent calculation proceeds as we compute the density matrix of the "extended molecule" $\rho_{ij}$ by integrating the "lesser" Green's function $G_{Dev}^<$ over a contour in the complex energy plane [20]

$$\rho_{ij} = \int \frac{dZ}{2\pi i} G_{Dev}^<(Z) \qquad (2)$$

Once the self-consistent calculation converges, all the physical observables can be obtained from the matrix Green's functions. In particular, the terminal current can be related to the Landauer formula through the transmission coefficient as:

$$I = \frac{2e}{h} \int_{-\infty}^{+\infty} T(E,V)\left[f(E-\mu_L) - f(E-\mu_R)\right] dE, \quad (3)$$

where the transmission coefficient is:
$$T(E,V) = Tr[\Gamma_L(E,V)G_{Dev}^r(E,V)\Gamma_R(E,V)G_{Dev}^a(E,V)], \quad (4)$$
and $\mu_R - \mu_L = eV$. The charge density is calculated from $\rho(\mathbf{r}) = \sum_{ij} \rho_{ij} \phi_i(\mathbf{r}) \phi_j^*(\mathbf{r})$. The local density of states (LDOS) is calculated using

$$\rho(r, E) = -\frac{1}{\pi} \lim_{\delta \to 0^+} \sum_{ij} \text{Im}[G_{ij}^r(E + i\delta)] \phi_i(\mathbf{r}) \phi_j^*(\mathbf{r})$$
$$= \sum_{ij} \rho_{ij}(E) \phi_i(\mathbf{r}) \phi_j^*(\mathbf{r}) \quad (5)$$

### 2.4. DENSITY FUNCTIONAL THEORY OF QUANTUM TRANSPORT?

Describing electron-electron interaction in single-molecule devices within the Kohn-Sham density functional theory simplifies tremendously the task of including electronic structure effect into the NEGF approach to quantum transport calculation. However, considering the ground state (or equilibrium state) nature of DFT, this approximation can only be considered as *ad hoc*. For example, the current computed using Eq. (3) corresponds to that carried by the (fictitious) Kohn-Sham orbitals of the "extended molecule". The relation between such computed current and current carried by the many-electron state of the "extended molecule" is not clear. Consequently, results of transport calculation based on the DFT theory of electron-electron interaction can only be taken qualitatively. Several recent works have proposed Non-Equilibrium Green's Function approach based on time-dependent density functional theory [36]. However, we are not convinced by such a approach since it neglects effect of the electromagnetic environment describing the external circuit.

The numerical results shown in this paper are obtained using a "minimal" microscopic model for implementing the "extended molecule" concept [25], and we will emphasize the chemical trends extracted from such computations whenever possible. To summarize, the metal electrodes are modeled as semi-infinite single-crystals. Six or seven (depending on the molecular adsorption site) nearest-neighbor metal atoms on each metal surface are included into the "extended molecule" where accurate first-principles based self-consistent calculation is performed. The rest of the electrodes (with the six/seven atoms on each surface removed) are considered as infinite electron reservoirs, whose effects are included as self-energy operators using tight-binding method. The calculation is performed using the Becke-Perdew-Wang parameterization of density-functional theory within the generalized-gradient approximation (GGA) [37] and appropriate pseudopotentials with corresponding optimized Gaussian basis sets [38].

## 3. What is Measured When You Measure the Conductance of a Molecule: The Effect of Metal-Molecule Interface

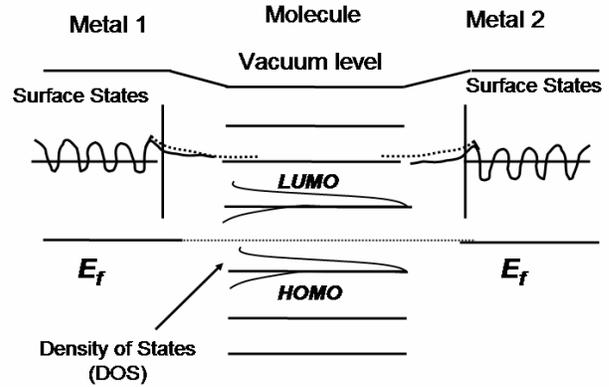

**FIGURE 2.** Schematic illustration of "Band" diagram (energy-level lineup) in two-terminal single-molecule device at zero bias. $E_f$ denotes the equilibrium metal Fermi-level. Upon contact to the electrodes, the frontier molecular states will be broadened, which is caused by the decay of the metal surface states into the molecular region and is characterized by the finite width in the density of states (DOS) plot.

The conductance of the molecular junction is an equilibrium property determined solely by the electronic processes in the equilibrium electrode-molecule-electrode junction, which can be computed using a number of existing electronic structure methods. The key electronic processes here are the charge transfer across the metal-molecule interface and the corresponding electrostatic potential change within the molecular junction, which modify the charge distribution of the molecular states and lineup their energy levels relative to the metal Fermi-level. Usually only several molecular states energetically closest to the metal Fermi-level (the so-called frontier molecular states) are relevant for the conduction process, including the highest-occupied-molecular-orbital (HOMO) and the lowest-unoccupied-molecular-orbital (LUMO).

The energy-level lineup is illustrated in Figure 2. Besides energy shift, the frontier molecular levels are also broadened, which can be analyzed through the density of states (DOS) of molecular junction, including both the total DOS and its projection onto individual molecular orbitals (PDOS). The broadening can be understood from the molecule perspective since the electrons residing on the molecular orbitals now have finite probability of escaping into the contacts due to the interface coupling. It can also be understood from the metal perspective. The electrons injected from the metal bulk are in evanescent states: their wavefuctions decay rapidly outside the metal surface. Inserting molecules in between the two electrodes slows down the decay of metal surface states or enhances the tails of metal surface states inside the molecule. This leads to additional density of states within the gap between initially discrete molecular levels. Since the total density of states is conserved, the metal-induced DOS has to be extracted from either the occupied molecular state or the unoccupied molecular states or both, which gives rise to the broadening of the individual molecular levels. Such metal-induced states have been called metal-induced gap states (MIGS) in the literature on metal-semiconductor interfaces [39]. Here we will call them instead as molecule-engineered surface states (MESS) to emphasize the active role molecules play in controlling the spatial character of the metal surface states propagating through the molecule.

Compared with other nanostructured devices, single-molecule device often requires attaching appropriate end groups chemically different from the molecule core in order to establish stable contact to the metal electrodes. Here we shall use as examples the prototypical molecular devices formed by attaching the benzene-based molecules onto gold (Au) electrodes through the sulfur end atoms, where the molecule bulk is composed of one to three benzene rings. The device structure adopted and the frontier molecular states of the molecules involved are shown in Figure 3. We assume that the molecules form symmetric contact with two semi-infinite gold <111> electrodes. The adsorption geometry is such that the molecule sits on-top of the triangular gold pad with an end sulfur-gold surface distance of 1.9 $\overset{\circ}{A}$. For the molecules with two and three benzene rings (BPD and TPD), a torsion angle of ~36 degrees exists between the neighbor benzenes, which reduces the corresponding orbital overlap and leads to an effective potential barrier for electron motion inside the molecule.

The introduction of end group (sulfur here) into the molecular structure has two immediate consequences: (1) it introduces molecular states that are end group-based; (2) it modifies the metal-molecule interaction through the metal surface-end group bond. For the phenyl dithiol molecules considered here, we find that the end group based states mainly belong to the occupied molecular states. Since the sulfur $p_\pi$ orbital can hybridize effectively with the $p_\pi$ orbital of internal carbon atoms, the introduction of sulfur end atoms lead to both localized and delocalized (occupied) molecular states.

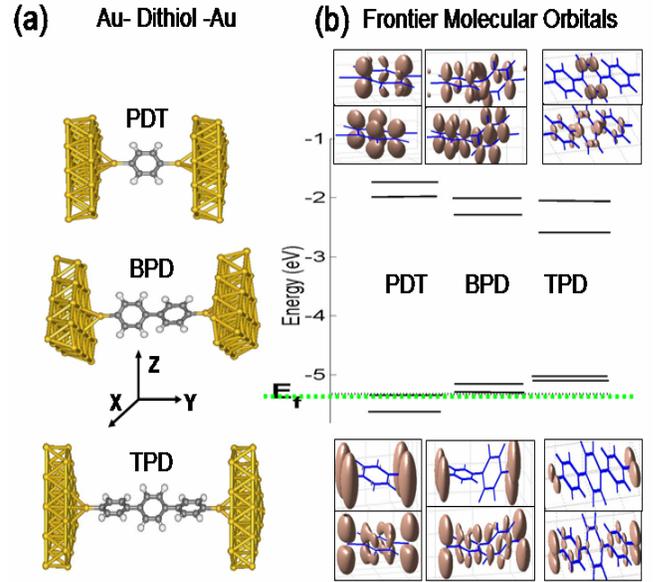

**FIGURE 3.** (a) Atomic structure of the gold-dithiol molecule-gold junction. The molecules considered are phenyl dithiol (PDT), biphenyl dithiol (BPD) and terphenyl dithiol (TPD) with one, two and three benzene rings respectively. (b) Orbital shape and energy of the four frontier molecular orbitals of the PDT, BPD and TPD molecules. The horizontal dotted line shows the location of the gold Fermi level $E_f$ at -5.31 eV.

The charge transfer upon junction formation is a localized process determined mainly by the local bonding across the metal-molecule interface. This is shown clearly in Figure 4(a). Since the PDT and TPD molecules have nearly identical interface configuration due to the identical end group and contact geometry, both the magnitude and the spatial distribution of the transferred charge are similar within the two molecular junctions. Due to the long-range Coulomb interaction between electrons, the electrostatic potential change within the molecular junction is determined by the charge density distribution throughout the molecular junction and is different both in its magnitude and spatial distribution for the two molecular junctions. The spatial distribution of the electrostatic potential change determines the lineup of the energy levels of the frontier molecular states relative to the metal Fermi-level.

Given the energy-level lineup of the frontier molecular orbitals and their broadening, the conductance of the molecule can be understood qualitatively from the two-level model considering only HOMO and LUMO-mediated transport, which gives the transmission coefficient through the metal-molecule-metal junction as:

$$T = \sum_{\substack{i=HOMO,\\LUMO}} \frac{\Gamma_{i;L}\Gamma_{i;R}}{(E_f - E_i)^2 + (\Gamma_{i;L} + \Gamma_{i;R})^2/4} \quad (6)$$

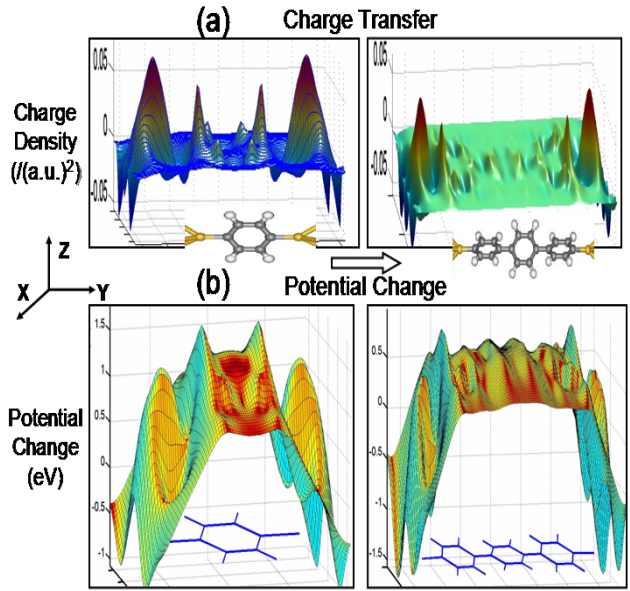

**FIGURE 4.** Charge transfer (a) and charge transfer induced electrostatic potential change (b) as a function of position in the XY-plane of the gold-PDT-gold and gold-TPD-gold junctions. The charge transfer (electrostatic potential change) is obtained from the charge density (electrostatic potential energy) difference between the molecular junction and the bare molecule *plus* the bare bi-metallic junction. We have also shown the coordinate system used. For the PDT molecule, the XY-plane is determined by the benzene ring while for the TPD molecule, it is determined by the left-most and right-most benzene rings.

where $\Gamma_{i;L(R)}$ denotes the broadening due to contact with the left (right) electrode respectively. The conductance is related to the transmission coefficient at energies around the Fermi-level over a range determined by the thermal broadening through the temperature dependence in the electrode Fermi distributions, and we can separate the tunneling and thermal-activation contribution to the conductance as follows:

$$G = \frac{2e^2}{h} \int_{-\infty}^{+\infty} T(E)\left[-\frac{df}{dE}(E-E_F)\right]dE = G_{Tunnel} + G_{Thermo}, \quad (7)$$

$$G_{Tunnel} = \frac{2e^2}{h}T(E_F), \quad G_{Thermo} = G - G_{Tunnel}.$$

The molecular junction conductance therefore is determined by both the lineup of the frontier molecular states relative to the metal Fermi level and the strength of their coupling to the metal surface states.

Both the energy level lineup and the conductance of the molecular junction can be examined from the equilibrium transmission versus energy (TE) characteristics of the molecular junction. For the three phenyl dithiol molecules studied here, we find that the metal Fermi-level lines up closer to the HOMO than to the LUMO upon contact with two gold electrodes. The result for the gold-PDT-gold junction is shown in Figure 5(a). Similar results are found also for the gold-BPD-gold and gold-TPD-gold junctions. The result for the room temperature molecular junction conductance as a function of the junction length is shown in Figure 5(b). For the short junctions (with lengths < 2.0 nm) considered here, the tunneling contribution is dominant and we can expect that the temperature dependence of the junction transport characteristics is weak. The dependence of the tunneling conductance on the junction length is different from the exponential law expected from simple tunneling models. This is not surprising since the sulfur end atoms introduce end group based states (Figure 3) which can be delocalized over a substantial part of the molecular junction. In general, we can expect a perfect exponential dependence of tunneling conductance on junction length only for molecular junctions long enough that the interface-induced perturbation of molecular states is suppressed completely in the interior of the molecule. The exact length where this occurs depends on both the end groups used and the internal structure of the molecules. For example, we have found that for saturated molecules like alkane dithiols, the exponential dependence of tunneling conductance on junction length is achieved at much shorter length than other molecules where the end group leads to delocalized molecular states (for example benzene molecules with cyanide end groups).

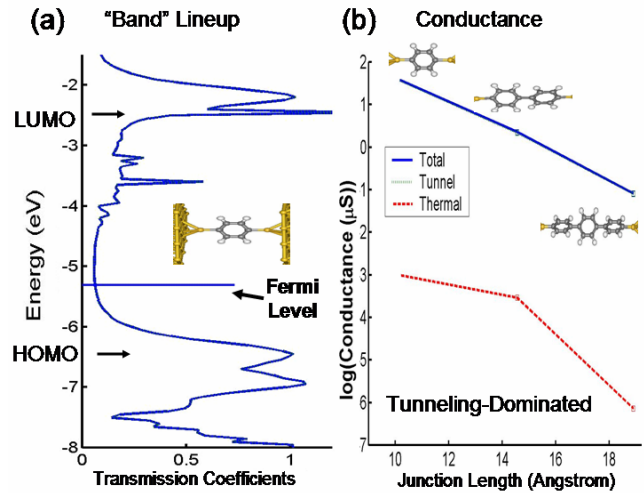

**FIGURE 5.** (a) Equilibrium transmission versus energy (TE) characteristics of the gold-PDT-gold junction. The first peak in the TE curve above the metal Fermi-level is due to tunneling through LUMO, while the first peak in the TE curve below the metal Fermi-level is due to tunneling through HOMO. (b) Room-temperature conductance of the gold-dithiol-gold junction as a function of the junction length. Here we have separated the thermal-activation contribution and the tunneling contribution. For short molecular junctions, the conductance is tunneling dominated (the tunneling contribution is virtually identical to the total conductance in the plot).

# 4. What is Measured When You Measure the Current Flowing Through a Molecule: The Local Field Effect

A finite current flows through the molecule as we drive the system out of equilibrium by applying a non-zero bias voltage across the two contacts through the external circuit (the battery). The bias voltage shifts the electrochemical potential deep inside the electrodes relative to each other. It also shifts the electrostatic potential deep inside the electrodes relative to each other, in order to maintain the thermodynamical equilibrium there [20, 22, 25]. The change in the electrostatic potential across the metal-molecule-metal junction will shift the molecular levels up or down relative to the two contact Fermi-levels. From Eq. (3), we can expect a significant increase in the total terminal current and correspondingly a peak in the conductance versus voltage characteristics, as one of the molecular levels is moved inside the energy window determined by the two contact Fermi-levels. This leads to the qualitative picture shown in Figure 6.

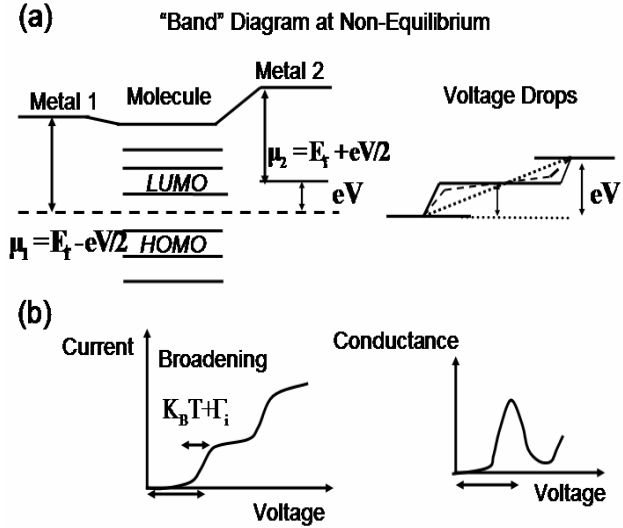

**FIGURE 6.** (a) Schematic illustration of "Band" diagram in two-terminal single-molecule device (left) and the electrostatic potential distribution across the molecular junction (right) at non-zero bias voltage V. Choosing the electrostatic potential energy in the middle of the bi-metallic junction as the energy reference, the electrochemical potential of the two electrodes is fixed by the applied bias voltage V at $\mu_1 = E_f - eV/2$ and $\mu_2 = E_f + eV/2$ respectively. Here the bias polarity is chosen such that for positive bias the electron is injected from the right electrode. (b) Schematic plot of the qualitative current- (left) and conductance- (right) voltage characteristics of the single-molecule device. There is a sharp increase in the terminal current at bias voltage where one molecular level is shifted passing one of the metal Fermi-levels. The sharp increase may be smoothed over a finite voltage range determined by the thermal (due to the electrode temperature) and electronic (due to the coupling to the electrodes) broadening. The conductance reaches a peak at the corresponding bias voltage.

We choose the electrostatic potential in the middle of bi-metallic junction (and in the region far away from the molecule) as the energy reference. The equilibrium Fermi level $E_f$ equals the negative of the metal work function (the average metal work function in the case of non-identical left and right electrodes), and the electrochemical potential of the two electrodes are fixed as $\mu_{1(2)} = E_f \mp eV/2$ respectively. The simplest physical picture of current flow assumes a voltage independent transmission coefficient, which at zero temperature gives $I = \frac{2e}{h} \int_{E_f - eV/2}^{E_f + eV/2} T(E) dE$. Using the two-level model of Eq. (6), the magnitude of the current is thus determined by the number of molecular levels falling within the energy window set by the contact electrochemical potentials $\mu_1 < E_i < \mu_2$, and the maximum current carried by one molecular level equals $I = \frac{2e}{\hbar} \frac{\Gamma_{i;L} \Gamma_{i;R}}{\Gamma_{i;L} + \Gamma_{i;R}}$. For a given total level-broadening by the two contacts $\Gamma_{i;L} + \Gamma_{i;R}$, the symmetric coupling case $\Gamma_{i;L} = \Gamma_{i;R}$ gives the largest current. Note that the rise to the current maximum is smoothed over a voltage range determined by the sum of the electronic broadening of $\Gamma_{i;L} + \Gamma_{i;R}$ and the thermal broadening of $k_B T$ at finite temperature.

Several important effects are neglected in this qualitative picture of non-equilibrium transport: (1) a finite electric field associated with the applied bias voltage will induce charge pileup (screening) within the molecular junction. The local transport field experienced by the tunneling electron is the self-consistent screened field [40]; (2) the local field can modify the charge distribution of the molecular states and their capability of carrying current [25]; (3) since the total current is conserved throughout the molecular junction for dc-transport, the spatial distribution of the current *density* can be highly non-uniform since the molecules are intrinsically inhomogeneous; (4) as current flows, the molecule is in general driven into a highly non-equilibrium state. It is not clear if an effective local electrochemical potential (LEP) can be defined everywhere within the molecular junction to characterize the electron distribution in the non-linear transport regime [41]. Indeed, answer to the above questions from the real-space view of current transport provides the theoretical basis for atomic engineering of single-molecule device operation. This can be analyzed through devices with different metal-molecule coupling and molecules with different internal structures.

The screened nature of the local transport field can be analyzed through field-induced charge redistribution inside the molecular junction and the resulting spatial profile of the local electrostatic potential. In particular, resistivity dipoles and strong local fields have been found in the vicinity of potential barriers that may exist both at the metal-molecule interface and inside the molecule [25]. These in general modify both the energy and wavefunction of the frontier molecular states and lead to different electrostatic potential drops depending on the molecule and the device structure. The resistivity dipole is a well known concept in mesoscopic transport [40], where their presence in the vicinity of local scattering center helps to overcome the barrier for transport and ensure current continuity. This can lead to a nonlinear transport effect due to the strong spatial variations in local carrier density and transport field [24].

The non-equilibrium charge pileup and the resulting voltage drop for the PDT molecule and the biphenyl dithiol (BPD) molecule with symmetric contact with two gold electrodes are shown in Figure 7(a). In the case of the PDT molecule, we can also create an asymmetric contact simply by increasing the right metal-molecule distance relative to the left metal-molecule interface. For the reference PDT molecule with symmetric contacts where the barrier for electron injection is located at the two gold-molecule interfaces, we found that both the resistivity dipole and the majority of the voltage drop are located (symmetrically) across the two interfaces. As we increase the right metal-molecule distance, a large barrier is formed for electron injected from the right contact. At positive bias voltage, we can create a strongly asymmetric situation where most of the voltage drop is located across the right interface with correspondingly a large charge decrease on the right end of the molecule, since the injected electrons from the right electrodes hit immediately the large barrier at the corresponding interface. For the BPD molecule in symmetric contact with the gold electrodes, an additional potential barrier exists inside the molecule due to the reduced orbital overlap across the two non-planar benzene rings. This leads to an additional resistivity dipole developing across the inter-benzene region. Correspondingly a large part of the voltage is also dropped inside the molecule as shown in Fig. 7(a).

The different spatial profile of the electrostatic potential drop leads to different physical picture of the bias-induced molecular level shift, as shown in Figure 7(b). The molecular levels plotted are calculated by diagonalizing the molecular part of the self-consistent Hamiltonian matrix at each bias voltage. For the gold electrodes considered here, the self-energy due to the contact depends only weakly on energy around the equilibrium Fermi-level. So the voltage dependence plotted here gives a faithful indication of the bias dependence of level shift, which can alternatively be obtained from the peak positions in the projected DOS plot [25]. We found that for the PDT and BPD molecules with symmetric contacts, the molecular levels are nearly constant at low bias but begin to shift at larger bias. The bias voltage at which this occurs corresponds to the voltage where the transmission coefficient begins to deviate significantly from that at zero bias, indicating that the shift of molecular levels is accompanied by the change in their charge distribution and therefore their capability of carrying current. For the PDT molecule with larger right metal-molecule distance, the molecular level shift follows that of the electrochemical potential of the left electrode at positive bias voltage, but shows more complicated pattern for negative bias voltage due to the different voltage drop profile there, which has a larger effect on the occupied molecular states than the unoccupied molecular states [25].

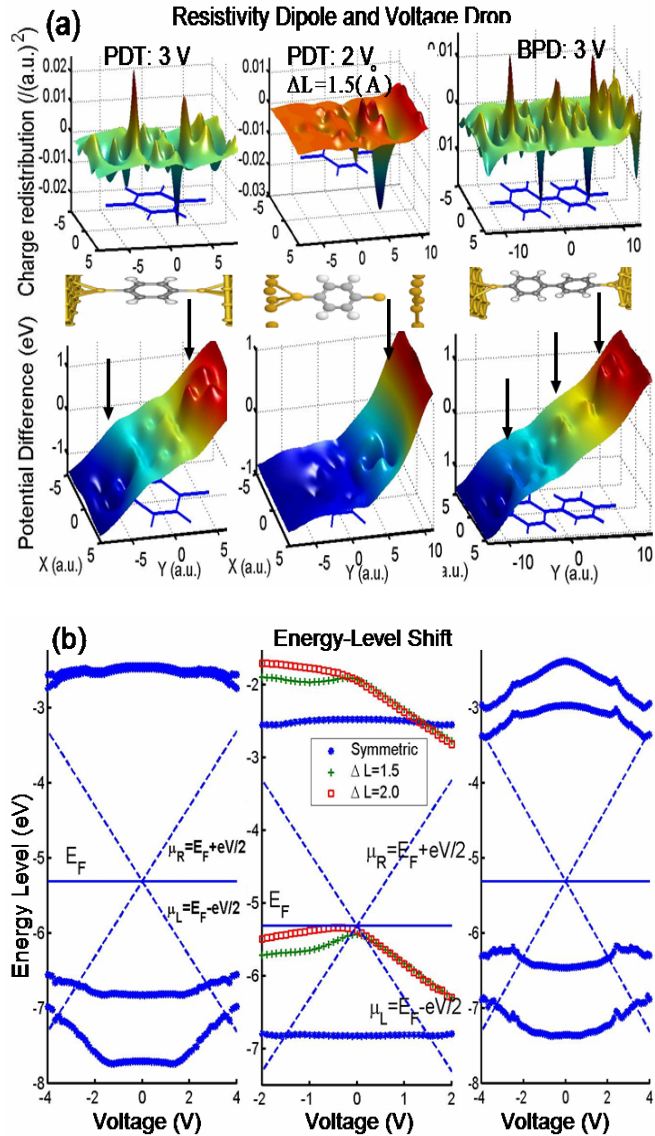

**FIGURE 7.** (a) Spatial distribution of charge transfer (upper figure) and potential drop (lower figure) at the gold-PDT-gold junction with symmetric contact, gold-PDT-gold junction with asymmetric contact ($\Delta L = 1.5 \text{ Å}$) and gold-BPD-gold junction with symmetric contact for bias voltages of 3.0 V, 2.0 V and 3.0 V respectively. Here $\Delta L$ denotes the difference between the right metal-molecule distance and the left metal-molecule distance. Also shown is the atomic structure of the molecular junctions. (b) Bias-induced molecular level shift for the three molecular junctions shown in (a).

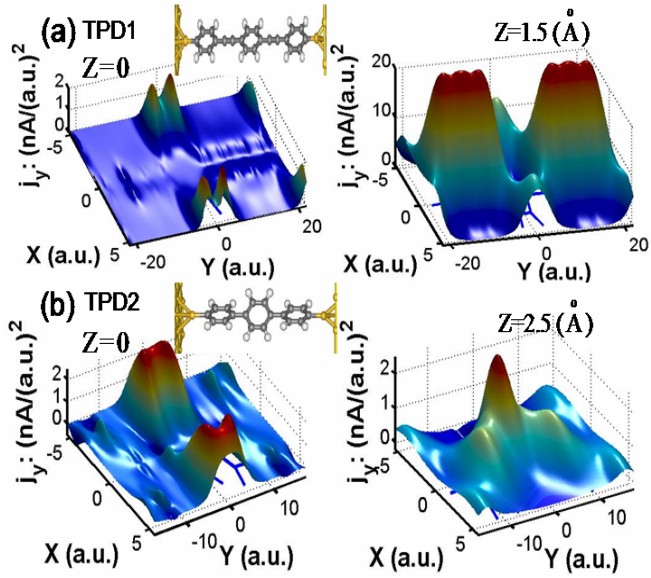

**FIGURE 8.** Cross sectional view of current density profiles at gold-TPD1-gold (a) and gold-TPD2-gold (b) junctions for bias voltage of 2 V. For each junction, we show the current density component along the transport direction (Y axis here). For the TPD1 junction, this is shown as a function of position in both the XY-plane and the plane that is located 1.5 Angstrom above. For the TPD2 junction, this is shown as a function of position in both the XY-plane and the plane that is located 2.5 Angstrom above. We have also shown the atomic structure of the molecules.

Further insight on current transport in single-molecule device can be obtained by analyzing in real-space the current density and local non-equilibrium electron distribution throughout the molecular junction. Since the effect is seen more clearly for longer molecules with internal barrier, we consider devices formed by a three-ring oligomer of phenyl ethynylene (called TPD1) and a terphenyl molecule (called TPD2)– in contact with gold electrodes through the end sulfur atoms, as shown in Figure 8. The internal geometry of the two molecules is different since the benzene rings are co-planar for the TPD1 molecule leading to optimal orbital overlap, but for the TPD2 molecule, a torsion angle is induced between neighbor benzene rings leading to weaker orbital overlap and an effective potential barrier for electron motion inside the molecule.

The overall current- and conductance- voltage characteristics of the two molecules show similar behavior. But the spatial distributions of current density are quite different due to the different internal structure of the molecules. For the planar TPD1 molecule, we show the y-component (perpendicular to the electrode surface) of current density $j_y$ in the XY-plane (defined by the benzene rings) and in the plane located 1.5 $\overset{0}{\text{Å}}$ above in Figure 8(a). For the non-planar TPD2 molecule, this is shown in the XY-plane (defined by the left-most and right-most benzene rings) and in the plane located 2.5 $\overset{0}{\text{Å}}$ above in Figure 8(b). In general we find that the current density distribution peaks at the same location where the delocalized pi-electron densities reach the maximum, and the direction of the current flow follows the internal structure of the molecule [42]. This is shown clearly in Figure 9(b) for the non-planar TDP2 molecule, where we find that the current density in the middle of the molecule is diverted toward the negative (positive) X half of the molecule at positive (negative) Z, following the location of carbon and hydrogen atoms in the central benzene ring.

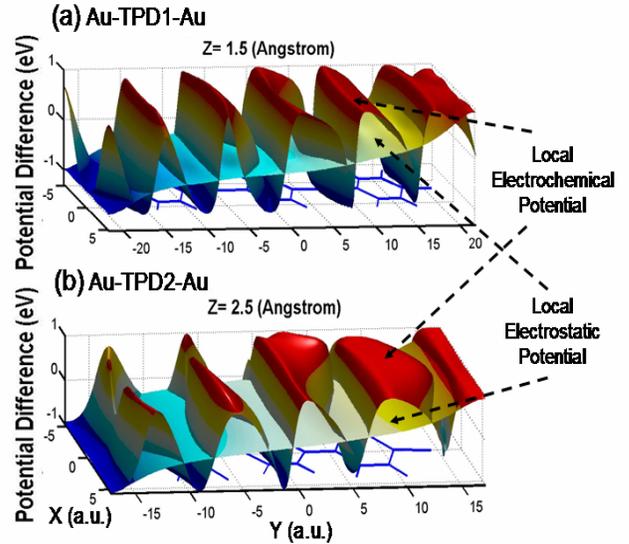

**FIGURE 9.** Cross sectional view of local electrostatic potential and electrochemical potential drop across the gold-TPD1-gold (a) and gold-TPD2-gold (b) junctions. Also shown is the position of the molecules.

The local electrochemical potential (LEP) characterizes the energy distribution of the electrons everywhere within the junction. For a coherent transport system, this can be defined effectively as [42,43]:

$$\mu_{eff}(\mathbf{r}) = \frac{\sum_{ij}(\rho_{L;ij}\mu_L + \rho_{R;ij}\mu_R)\phi_i(\mathbf{r})\phi_j(\mathbf{r})}{\sum_{ij}\rho_{ij}\phi_i(\mathbf{r})\phi_j(\mathbf{r})} \quad (8)$$

where $\phi_i(\mathbf{r})$ is the atomic basis function. Note that $\rho_{ij} = \rho_{L;ij} + \rho_{R;ij}$ is the $ij$-th element of the local density of states at non-equilibrium as defined in Eq. (5), which can be decomposed into contact-resolved contributions as $\rho_{L(R)} = G^r \Gamma_{L(R)} G^a / 2\pi$. The contact-resolved density of states $\rho_{L(R)}$ is proportional to the injectivity of the left (right) contact given by Gramespacher and Buttiker [43], who gives an equivalent definition of the local electrochemical potential starting from the scattering matrix theory. The LEP as defined here may possibly be measured using an ideal phase-sensitive non-invasive voltage probe (Buttiker probe) [43]. In scanning tunneling potentiometry (STP), in principle this can be identified with the bias voltage between the STP tip and the sample under the condition of zero tunneling current [41].

The local electrostatic potential and electrochemical potential drops in the two molecular junctions are shown in Figure 9 for bias voltage of 2 V. The electrostatic potential drop varies rather smoothly from -V/2 to V/2 across the junction, but the local electrochemical potential shows oscillatory behavior between -V/2 and V/2 throughout the molecular junction due to the phase-sensitive nature of the voltage probe: an electron wave incident from either electrode gives two contributions to the electron flux in the voltage probe at point $\mathbf{r}$ through the direct transmission and the transmission after multiple scattering within the molecular junction. In practice, the strong local variation of LEP may be extremely difficult to observe due to the spatial resolution involved (half of the metal Fermi wavelength or $\sim 1 \ \overset{\circ}{\text{A}}$) and the requirement that the local probe should be minimally invasive. Recently scanning probe measurement of local voltage drop across a metal-carbon nanotube-metal junction has been reported by Yaish *et al* [44]. However, due to the large spatial resolution (tens of nm) there, the measured electrochemical potential drop is the result of spatially-averaging over many oscillation periods of the local electrochemical potential, which essentially follows the local electrostatic potential drop.

## 5. Conclusion

In conclusion, we have presented a general theoretical framework for analyzing single-molecule electronics by extrapolating systematically the concepts and techniques of mesoscopic transport into the molecular regime. The real-space formulation of our approach allows us to give an intuitive and coherent physical picture of single-molecule electronics by establishing a clear connection between the transport characteristics and the molecular electronic structure perturbed by the metal-molecule coupling and the applied bias voltage.

Historically, physics of mesoscopic transport systems were developed in the context of disordered metals and artificial semiconductor microstructures fabricated through the top-down lithographic approach [45]. The advances in the synthetic bottom-up nanofabrication approach have in turn made molecular electronics an ideal laboratory that may help in ushering in new theoretical principles of nanoscopic physics. The field of molecular electronics therefore serves the dual purposes of exploring new device technologies through the atom-engineering technique and providing the testing ground of new physical principles and new theoretical and computational techniques. Such theoretical developments are currently under active investigation both in our groups and in other research groups. The work presented here therefore represents only a snapshot of where we stand in the fascinating field of molecular electronics.

## ACKNOWLEDGEMENT

This paper is dedicated to the memory of the late Professor John A. Pople, whose pioneering work on theoretical and computational chemistry has had a profound influence on our research in the field of molecular electronics. This work was supported by the DARPA Moletronics program, the DoD-DURINT program and the NSF Nanotechnology Initiative.